\PassOptionsToPackage{a4paper,left=1.5cm,right=1.5cm,top=1cm,bottom=1.5cm}{geometry}
\documentclass[a4paper,10pt,5p]{elsarticle}
\usepackage{amsmath}
\usepackage{amssymb}
\usepackage{graphicx}
\usepackage{rotating}
\usepackage[usenames,dvipsnames]{xcolor}
\usepackage[colorlinks=true,urlcolor=blue,linkcolor=RubineRed]{hyperref}
\usepackage{attachfile}

\attachfilesetup{color=0 0.5 0.5,print=false,mimetype=text/plain,icon=Paperclip,description=Save source file,author=M. Yu. Uleysky}
\newcommand{\att}[1]{\textattachfile{#1}{\protect\smash{\protect\includegraphics[width=0.75em]{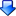}}}}
\newcommand{\natt}{\notextattachfile[print=true]{\protect\smash{\protect\includegraphics[width=0.75em]{kget.png}}}}
\newcommand{\attv}[1]{\textattachfile{#1}{\protect\smash{\protect\includegraphics[height=0.96em]{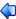}}}}

\begin{document}
\begin{frontmatter}
\title{Genealogical tree of Russian schools on Nonlinear Dynamics}
\author{S. V.~Prants}
\address{Laboratory of Nonlinear Dynamical Systems,\\
Pacific Oceanological Institute of the Russian Academy of Sciences, \\
43 Baltiiskaya st., 690041 Vladivostok, Russia}
\ead{prants@poi.dvo.ru}
\ead[url]{http://dynalab.poi.dvo.ru}
\author{M. Yu.~Uleysky}
\begin{abstract}
One of the most prominent feature of research in Russia and the former Soviet Union
is so-called scientific schools. It is a collaboration of researchers with a common scientific background working, as a rule,
together in a specific city or even at an institution. The genealogical tree of scientific schools on nonlinear dynamics
in Russia and the former Soviet Union is grown. We use these terminology in a broad sense including theory of dynamical
systems and chaos and its applications in nonlinear physics. In most cases we connect two persons if one was an  advisor of the Doctoral thesis
of another one. It is an analogue of the Candidate of Science thesis in Russia.
If the person had no official advisor or we don't know exactly who was an advisor, we fix that person who was known to be an informal teacher and
has influenced on him/her very much.
\end{abstract}
\end{frontmatter}

\thispagestyle{plain}
\section{Introduction}

The intent of our project is to grow a genealogical tree of
nonlinear dynamics in Russia and the former Soviet Union. We use
these terminology in a broad sense including theory of dynamical
systems and chaos and its applications in nonlinear physics.
% (nonlinear optics, hydrodynamics, statistical physics, plasma physics, etc.)
%and physical oceanography.

In most cases we connect two persons if one was an  advisor of the Doctoral thesis
of another one. It is an analogue of the Candidate of Science thesis in Russia.
If the person had no official advisor or we don't know exactly
who was an advisor, we fix that person who was known to be an informal teacher and
has influenced on him/her very much.

One of the most prominent feature of research in Russia and the former Soviet Union
is so-called scientific schools. It is a
collaboration of researchers with a common scientific background working, as a rule,
together in a specific city or even at an institution. Mandelstamm's, Landau's,
Kolmogorov's, Tamm's, Andronov's are famous examples. We tried to specify a few schools
including the Moscow mathematical school on dynamical system theory and
the Nizhny Novgorod (former Gorky), Novosibirsk, Saratov and Vladivostok schools on nonlinear dynamics.

\begin{sidewaysfigure*}
\centerline{\includegraphics[width=0.9\textwidth]{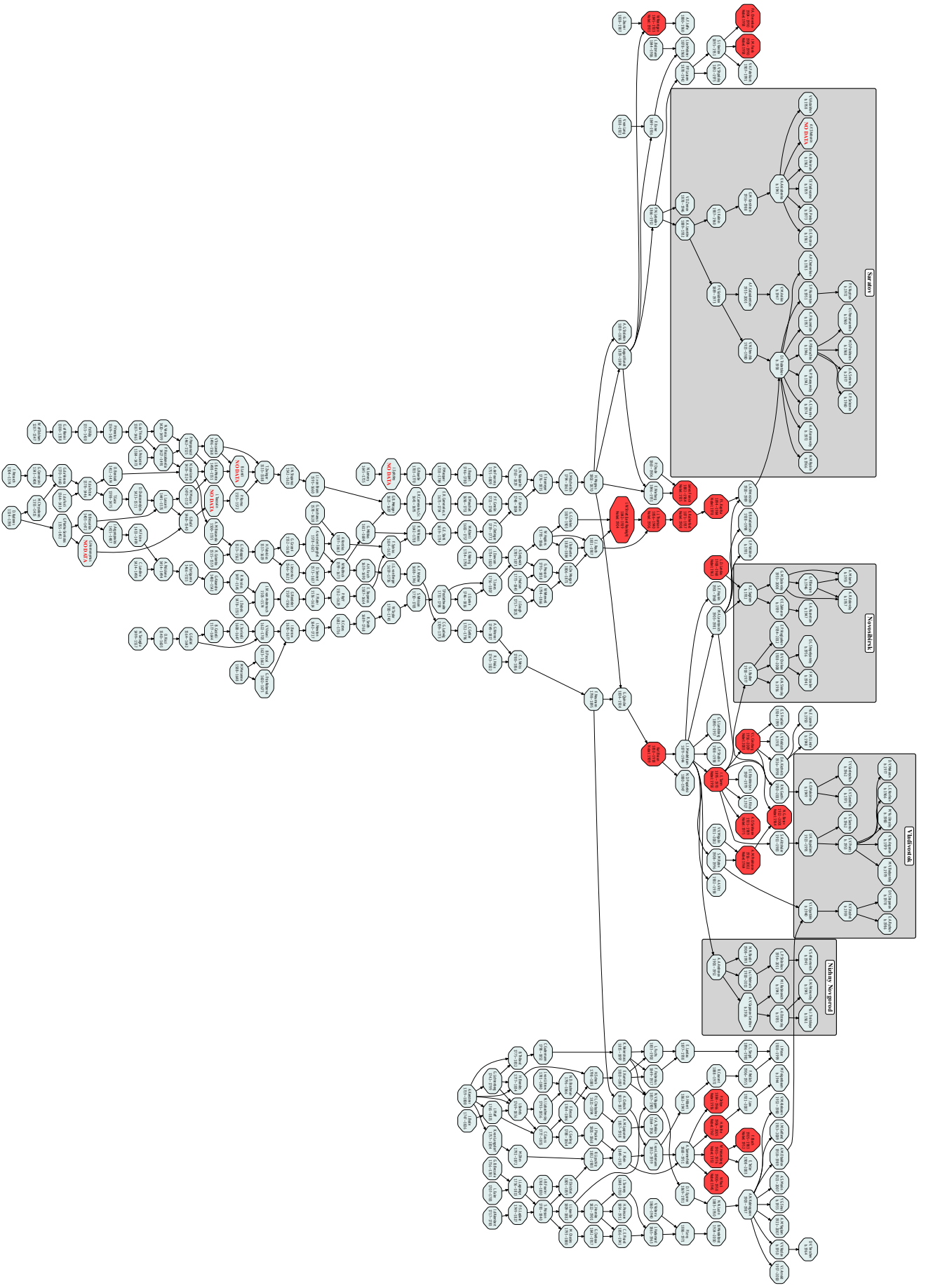}}
\caption{The Russian tree on Nonlinear Dynamics. \attv{gen.dot}}
\label{gen_en}
\end{sidewaysfigure*}

\begin{figure*}[!htp]
\centerline{\includegraphics[width=\textwidth]{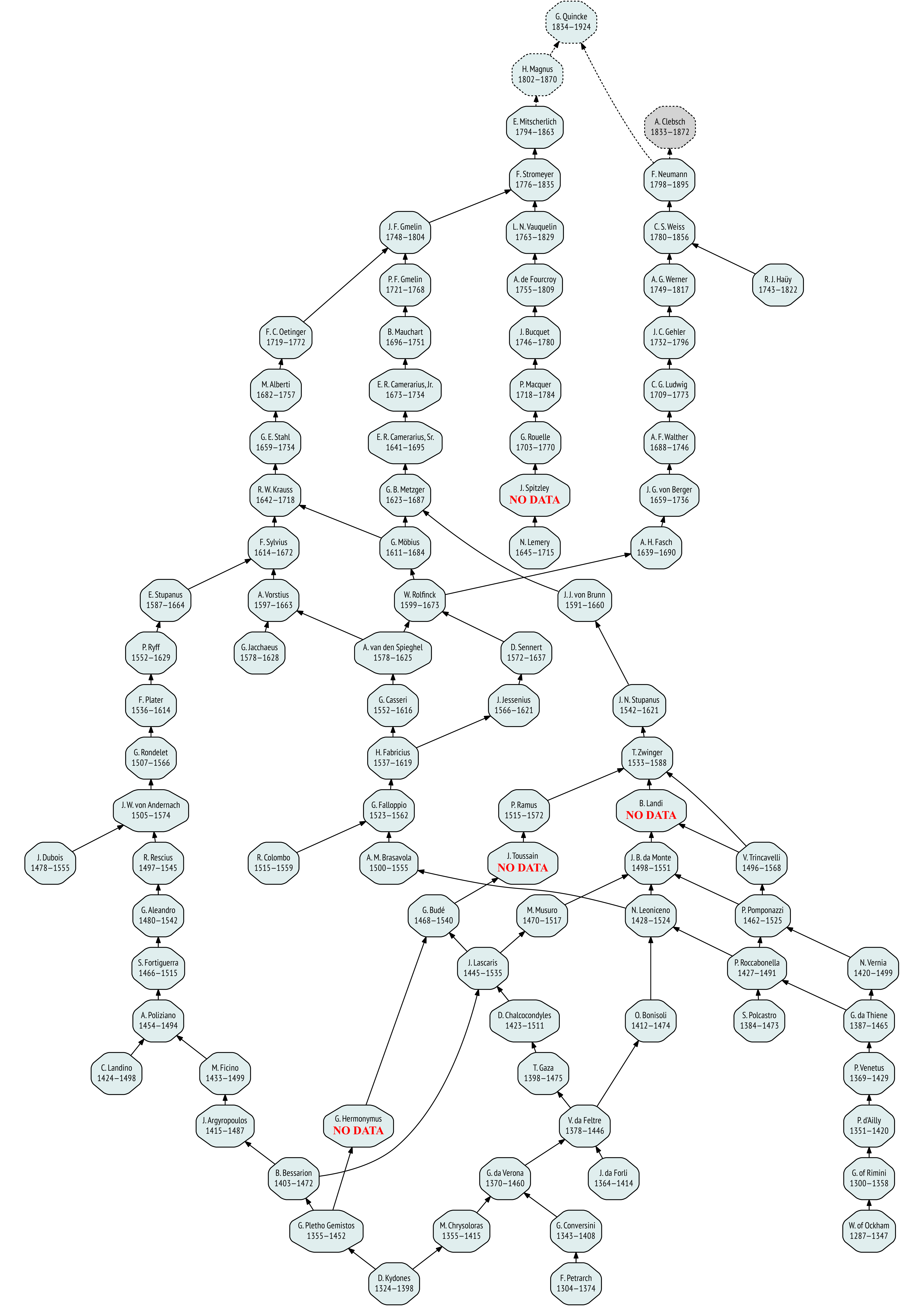}}
\caption{The British--German--Italian branch. \att{phys1.dot}}
\label{phys1_en}
\end{figure*}

\begin{sidewaysfigure*}[!htp]
\centerline{\includegraphics[width=\textwidth]{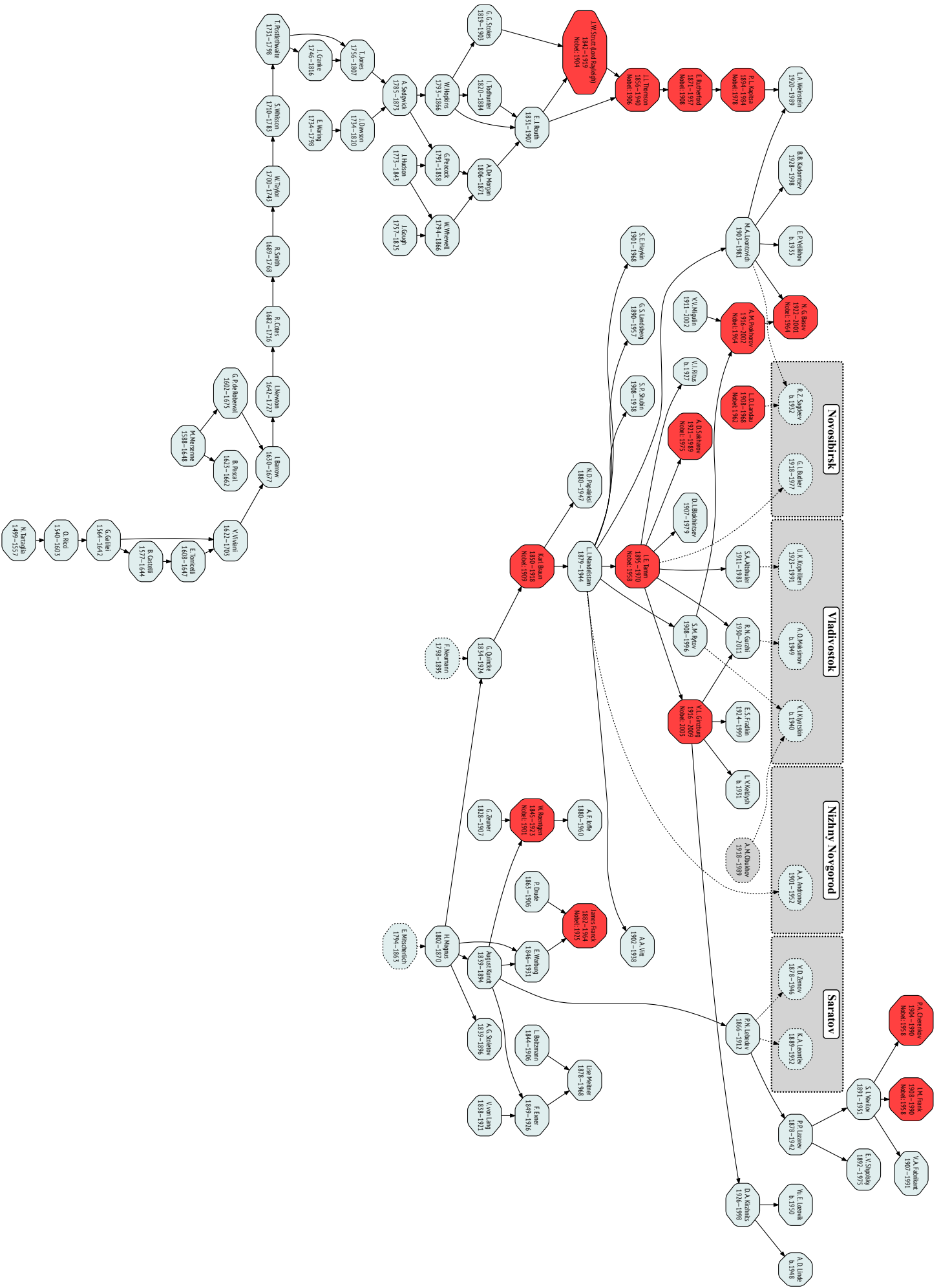}}
\caption{The Petrarch--Ockham branch. \attv{phys2.dot}}
\label{phys2_en}
\end{sidewaysfigure*}

For better viewing the material is structured as follows. In
Fig.~\ref{gen_en} we present a general image of the Russian tree on
nonlinear dynamics. It is a huge tree with the roots which we were
able to trace down to the thirteen century. The picture can be
zoomed using common Adobe Reader tools. Then we split the tree into
two parts and show them separately. The
first part in Fig.~\ref{phys1_en}, which we call the Petrarch--Ockham
branch, shows the roots which go down to medieval European
naturalists and philosophers. The second part in
Fig.~\ref{phys2_en} show British, German and Italian roots of the
Nizhny Novgorod, Novosibirsk, Saratov and Vladivostok schools which
go down to medieval Italian physicists and mathematicians. Each Russian scientific school is
presented in detail in a separate section and is shown on a graph.


\section{Moscow mathematical school on dynamical system theory}

The fruitful Kolmogorov's mathematical school at the Lomonosov State
University in Moscow is widely known for great achievements in
dynamical system theory and mechanics by A.~Kolmogorov (1903--1987)
and his former students, V.~Arnold (1937--2010), Ya.~Sinai,
I.~Gel'fand (1913--2009), V.~Alekseev (1932--1980), A.~Yaglom
(1921--2007), A.~Obukhov (1918--1989), A.~Monin (1921--2007) and many
others. N.~Luzin (1883--1950) was the Kolmogorov's doctoral advisor.
The branches from N. Luzin go via the Russian mathematicians
D.~Egorov (1869--1931) and N.~Bugaev (1837--1903) to the French
mathematicians, J.~Liouville (1809--1882), S.~Poisson (1781--1840),
P.-S.~Laplace (1749--1827), J.~Lagrange (1736--1813), J.~d'Alembert
(1717--1783) and to the German ones, K.~ Weierstrass (1815--1897),
E.~Kummer (1810--1893), F.~Bessel (1784--1846), C.~Gauss (1777--1855)
and L.~Euler (1707--1783). The other branch from N.~Bugaev goes down
to the German professor A.~Kaestner (1719--1800). One of his branches
goes upward to the mathematicians R.~Lipschitz (1832--1903) and
F.~Klein (1849--1925) and then to the German physicist A.~Sommerfeld
(1868--1951) and his famous school on quantum physics. On another
Kaestner's branch we meet the Russian mathematicians
N.~I.~Lobachevsky (1792--1856), P.~L.~Chebyshev (1821--1894) and
A.~M.~Lyapunov (1857--1918).
\begin{figure*}[!htp]
\centerline{\includegraphics[width=\textwidth]{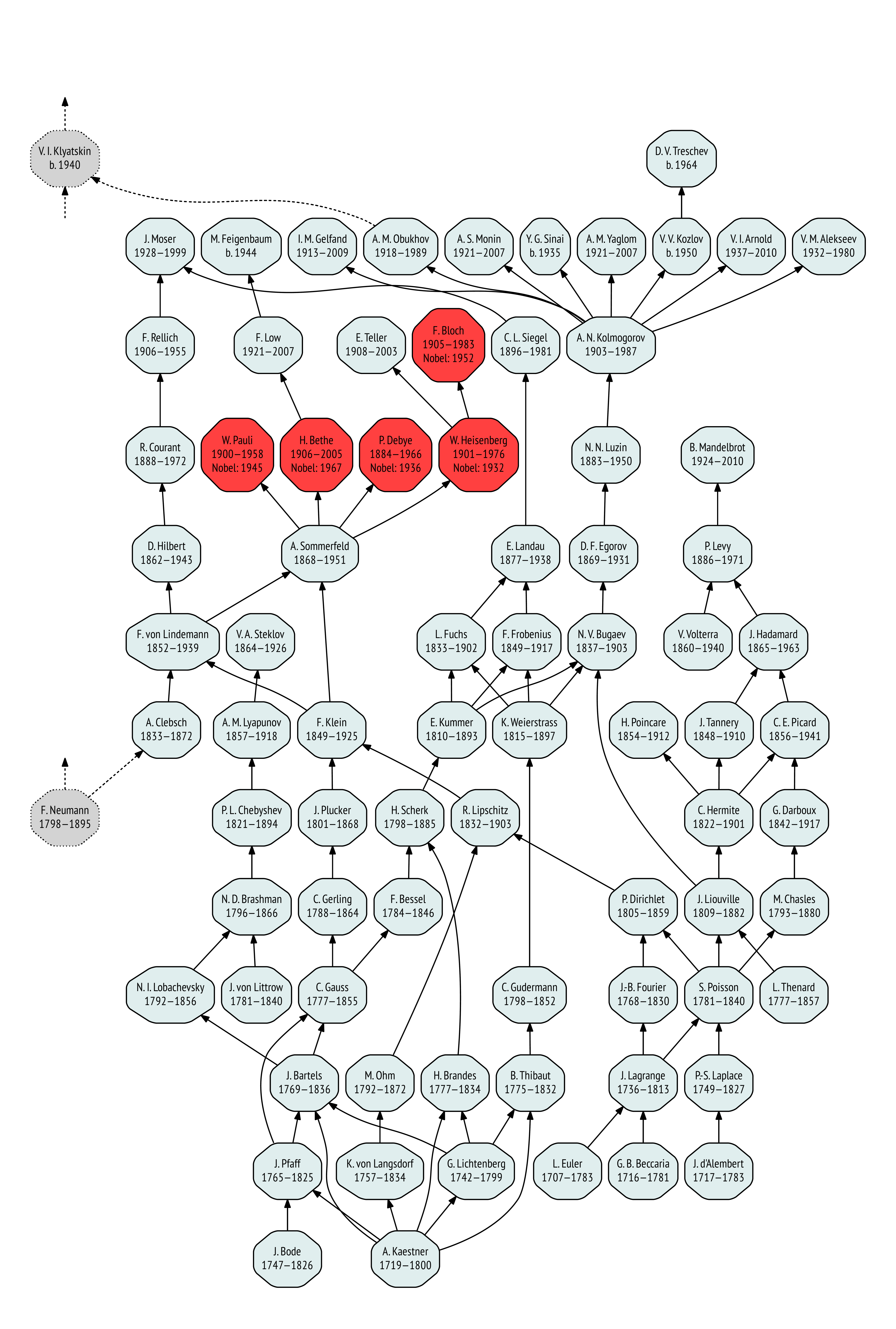}}
\caption{The Moscow mathematical school on dynamical system theory. \att{math.dot}}
\label{math_en}
\end{figure*}

\section{Nizhny Novgorod (former Gorky) school on nonlinear oscillations and dynamical system theory}

The Nizhny Novgorod (former Gorky) school on nonlinear oscillations and dynamical system theory
was founded by A.~Andronov (1901--1952) who was a student of  L.~ Mandelstam (1879--1944).
K.~ Braun (1850--1918), a German inventor and Nobel laureate in physics, was Mandelstam's doctoral advisor.
That branch via the German scientists
G.~Quincke (1834--1924), H.~Magnus (1802--1870) and F.~Neumann (1798--1895) goes, on one hand,
to the German physicists and mathematicians,
A.~Clebsch (1833--1872), D.~Hilbert (1862--1943), A.~Sommerfeld and others.
The mathematical branch from D.~Hilbert
goes upward via R.~Courant (1888--1972) and F.~Rellich  (1906--1955) to J.~Moser (1928--1999).
The other branch to J.~Moser is connected to A.~Kolmogorov
via K.~Weierstrass and E.~Kummer  (1810--1893) and N.~Bugaev. Thus, all the authors of the famous
Kolmogorov--Arnold--Moser
theorem turn out to be scientific relatives. A.~Sommerfeld was a doctoral advisor for a few
Nobel laureates in physics, P.~Debye (1884--1966), W.~Pauli (1900--1958), W.~Heisenberg (1901--1976),
F.~Bloch (1905--1983) and H.~Bethe (1906--2005).

On the other hand, the branch from G.~Quincke and F.~Neumann goes down to medieval
naturalists and to the great Italien poet F.~Petrarch (1304--1374).
The branch from G.~Quincke and H.~Magnus
goes down to medieval naturalists and to the English Franciscan friar and
philosopher W.~of~Ockham (1287--1347).
Italien mathematician and engineer N.~Tartaglia (1499--1557) from Republic of Venice
has been found to be connected via  G.~Galilei (1564--1642), V.~Viviani (1622--1703)
and I.~Barrow (1630--1677)
to I.~Newton (1642--1727). The Newton's branch goes upward to the British physicists
including G.~G.~Stokes (1819--1903) and  Nobel laureates in physics J.~W.~Strutt (Lord Rayleigh)
(1842--1919), J.~J.~Thomson (1856--1940) and
E.~Rutherford (1871--1937) who was an advisor of the Nobel laureate in physics P.~Kapitza (1894--1984).
\begin{figure}[!htp]
\centerline{\includegraphics[width=0.5\textwidth]{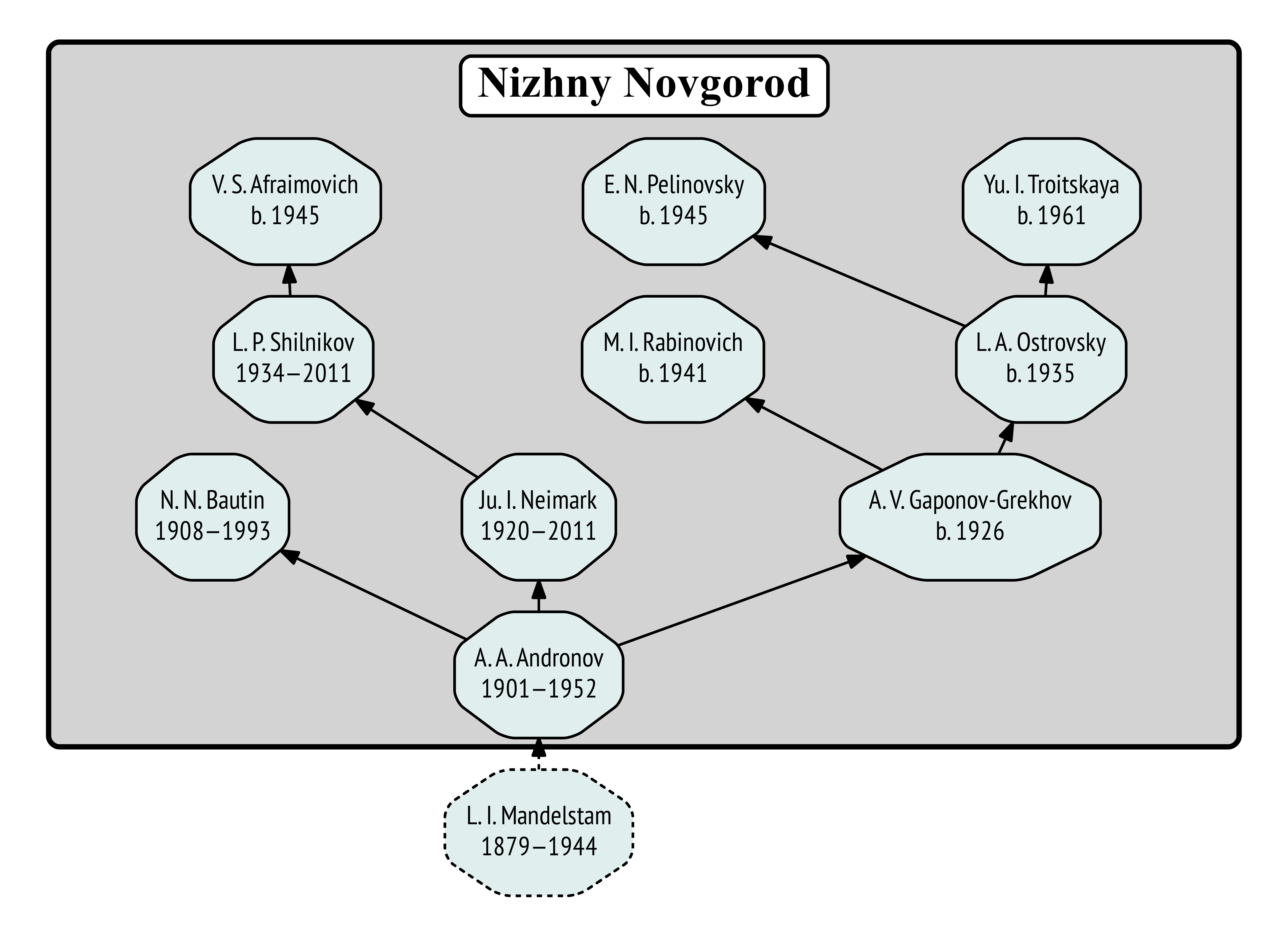}}
\caption{The Nizhny Novgorod (former Gorky) school on nonlinear oscillations and dynamical system theory. \att{Novgorod.dot}}
\label{Novgorod_en}
\end{figure}

\section{Saratov school on nonlinear oscillations and chaos}

One branch of the Saratov school on nonlinear oscillations and chaos goes from
P.~Kapitza to L.~A.~Weinstein (1920--1989) and then to D.~I.~Trubetskov who was a Doctoral advisor of
many modern physicists and mathematicians in Saratov. The second branch of the Saratov school
via K.~A.~Leont'ev (1889--1932) goes to the Russian physicist P.~N.~Lebedev (1866--1912),
the first person who measurered the light pressure. He was a student of
August Kundt (1839--1894) whose scientific roots are hidden in Middle Ages and
could be traced to W.~of~Ockham and F.~Petrarch.
It is interesting that the Saratov school via L.~A.~Weinstein and
M.~A.~Leontovich (1903--1981) is connected again with L.~I.~Mandelstam.
\begin{figure*}[!htp]
\centerline{\includegraphics[width=0.95\textwidth]{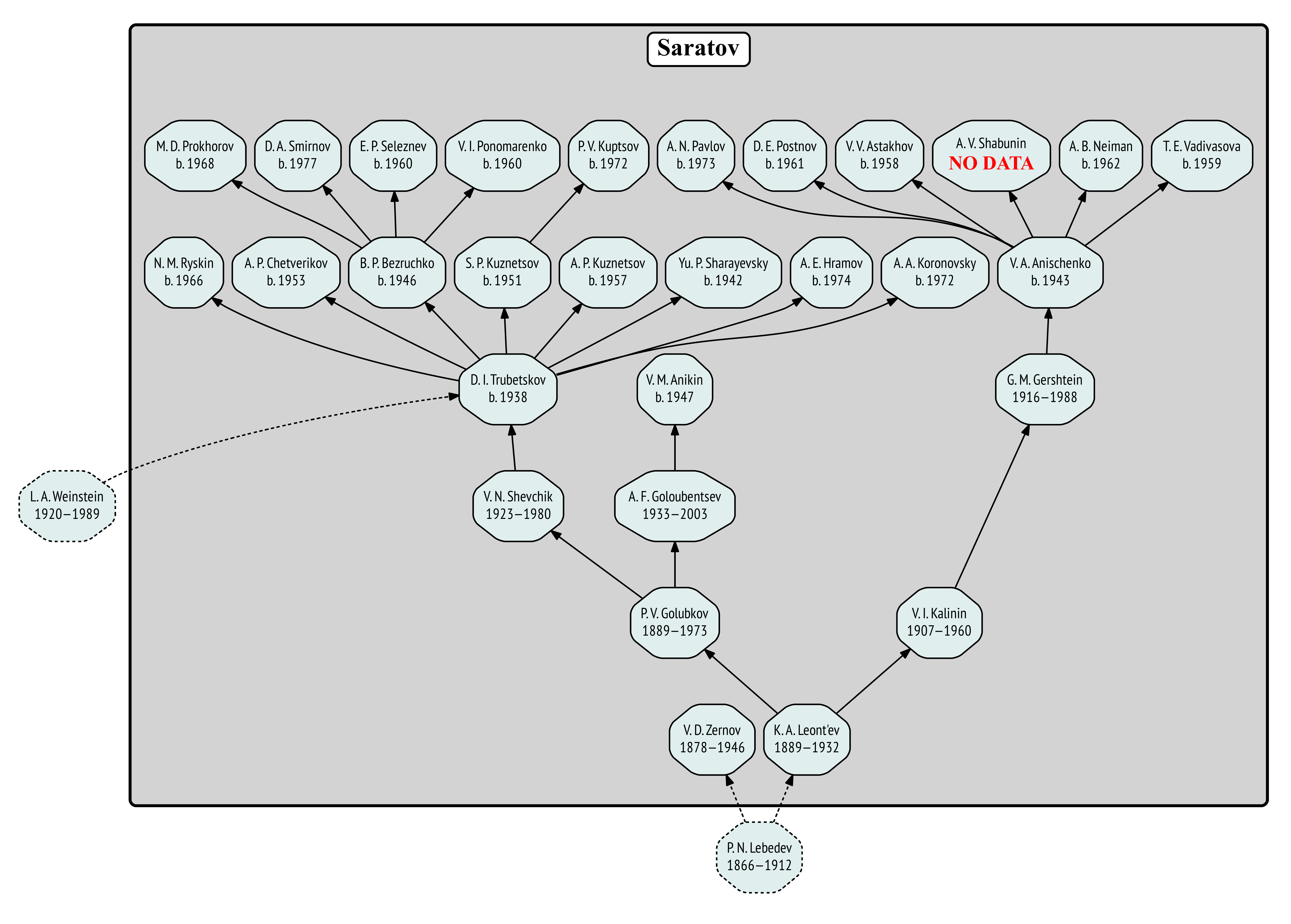}}
\caption{The Saratov school on nonlinear oscillations and chaos. \att{Saratov.dot}}
\label{Saratov_en}
\end{figure*}

\section{Novosibirsk school on nonlinear physics and chaos}

L.~I.~Mandelstam was also one of the founders of the Novosibirsk school on nonlinear physics and chaos.
One Mandelstam's branch goes to M.~A.~Leontovich, then to
R.~Z.~Sagdeev and his school on nonlinear physics including G.~M.~Zaslavsky (1935--2008)
and V.~E.~Zakharov each of which has his one school in Siberia, Moscow and the USA.
The other Mandelstam's branch goes to the Nobel laureate in physics I.~E.~Tamm (1895--1970)
and then to G.~I.~Budker (1918--1977), a founder of the Institute of Nuclear Physics
of the Siberian Branch of the Russian Academy of Sciences in Novosibirsk.
B.~V.~Chirikov (1928--2008), known as one of the key persons in theory of Hamiltonian chaos,
worked at that institute. We show only a small part of the  Novosibirsk school
which is much more extensive.
\begin{figure*}[!htp]
\centerline{\includegraphics[width=0.8\textwidth]{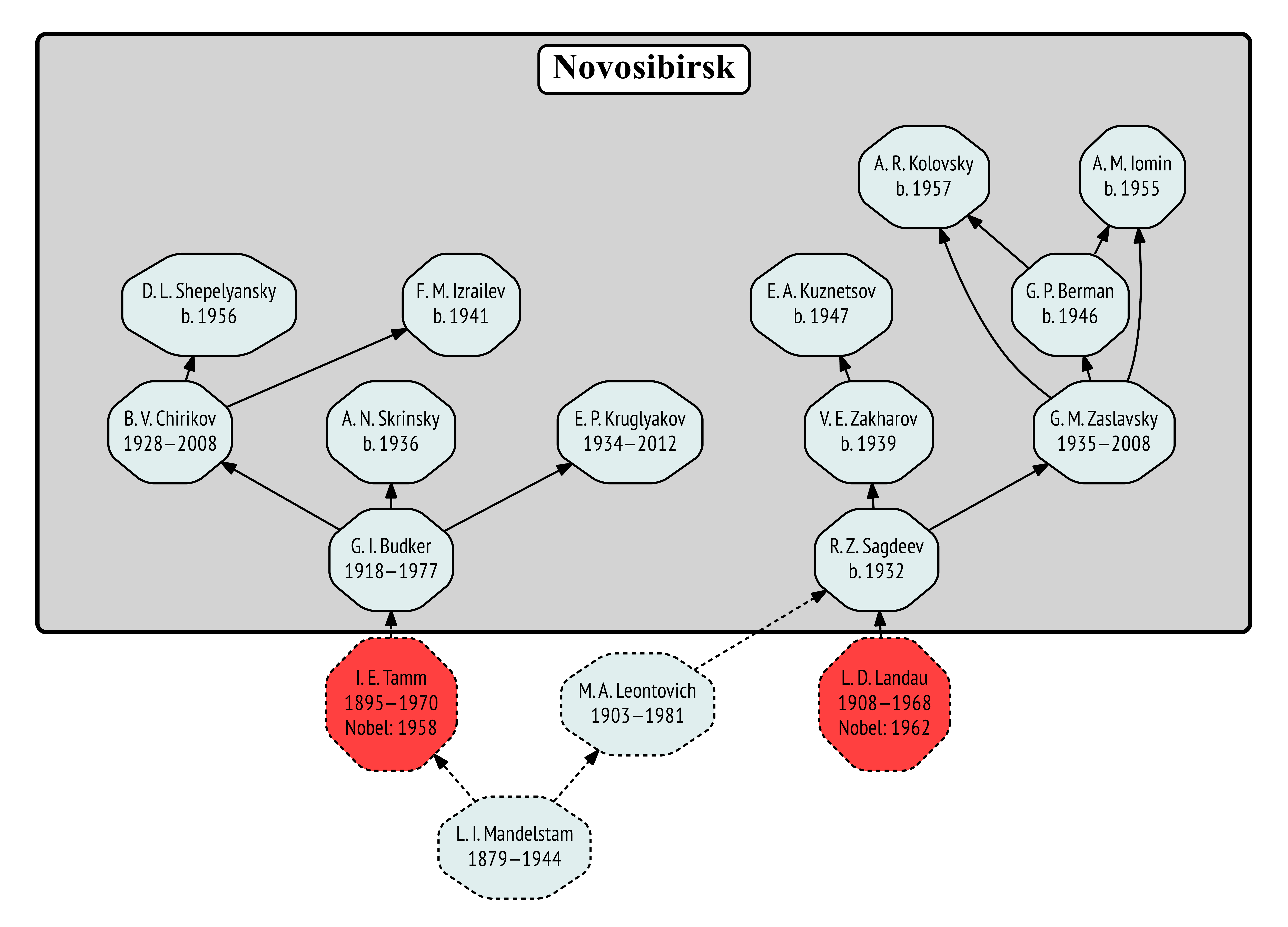}}
\caption{The Novosibirsk school on nonlinear physics and chaos. \att{Novosibirsk.dot}}
\label{Novosibirsk_en}
\end{figure*}

\section{Vladivostok school on nonlinear oceanography and chaos}

In fact, L.~I.~Mandelstam was a founder of all the Russian schools on
nonlinear dynamics and nonlinear physics including the most remote and young one in Vladivostok.
One Mandelstam's branch via I.~E.~Tamm, S.~A.~Altshuler (1911--1983) and U.~Kh.~Kopvillem
(1923--1991) forms the modern Vladivostok school on nonlinear oceanography and
dynamical system theory approch to study the ocean and atmosphere.
The other one via the Nobel laureates in physics I.~E.~Tamm and  V.~L.~Ginzburg
(1916--2009) and R.~N.~Gurzhi (1930--2011) forms now the Vladivostok school
on nonlinear oscillations and acoustics. The third Mandelstam's branch via S.~M.~Rytov (1908--1996) and
V.~I.~Klyatskin forms the modern Vladivostok school on hydrodynamical chaos and statistical physics.
This branch via  A.~M.~Obukhov (1918--1989) is connected also to
A.~N.~Kolmogorov's Moscow mathematical school
on dynamical system theory and mechanics and to many French and German mathematicians.
%as it is shown above.
%
\begin{figure*}[!htp]
\centerline{\includegraphics[width=\textwidth]{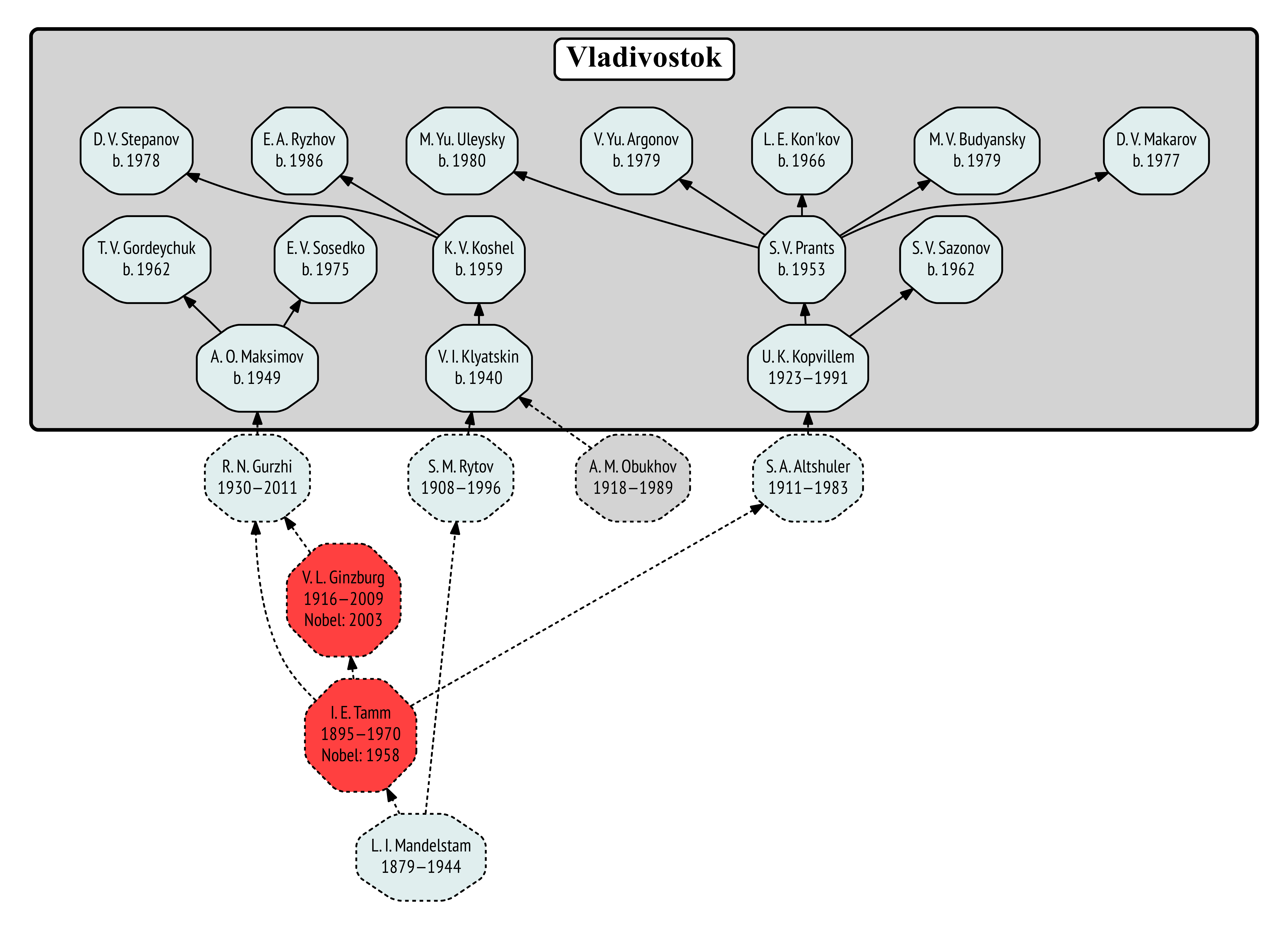}}
\caption{The Vladivostok school on nonlinear oceanography and chaos. \att{Vladivostok.dot}}
\label{Vlad_en}
\end{figure*}

Everybody is welcome to complete the tree and we would appreciate any corrections,
comments and additions.
The graphs were created by a \textbf{dot} program from the Graphviz package
(\mbox{\url{http://www.graphviz.org/}}).
The source files for each tree can be found by clicking on the arrow {\natt} after the
corresponding figure caption.

The following sources have been used:

\begin{itemize}
\item Mathematics Genealogy Project\\(\url{https://www.genealogy.ams.org}).
\item The Academic Family Tree\\ (\url{http://academictree.org/}).
\item Russian (\url{https://ru.wikipedia.org/wiki}) and English (\url{https://en.wikipedia.org/wiki}) Wikipedia.
\item Private communications. We thank V.~Afraimovich (San Luis Potosi, Mexico),
F.~Izrailev (Puebla, Mexico), V.~Klyatskin (Moscow, Russia), E.~Kuznetsov (Moscow, Russia),
S.~Kuznetsov (Saratov, Russia), A.~Maksimov (Vladivostok, Russia),
L.~Ostrovsky (Boulder, USA), D.~Shepelyansky (Toulouse, France),
V.~Sokolov (Novosibirsk, Russia).
\item Biographical and autobiographical texts.
\end{itemize}

\end{document}